\documentstyle[sprocl]{article}

\input{psfig}

\def\be{\begin{equation}}
\def\ee{\end{equation}}
\def\bea{\begin{eqnarray}}
\def\eea{\end{eqnarray}}

\def\to{\rightarrow}

\def\ttau{\tilde\tau}

\def\ttau{\tilde \tau}
\def\tg{\tilde g}

\def\tq{\tilde q}

\def\tw{\widetilde W}

\newcommand{ \ltap }{\stackrel{\lower.65ex\hbox{$<$}}
                     {\lower.65ex\hbox{$\sim$}}}
\newcommand{ \gtap }{\stackrel{\lower.65ex\hbox{$>$}}
                     {\lower.65ex\hbox{$\sim$}}}

\begin{document}

\title{$b\rightarrow s\gamma$ DECAY IN SUPERSYMMETRIC THEORIES}

\author{M. BRHLIK}

\address{Randall Physics Lab, University of Michigan\\
Ann Arbor, MI 48109-1120, USA\\E-mail: mbrhlik@umich.edu}


\maketitle\abstracts{ 
Recent advances in the calculation of the $b\rightarrow s\gamma$ decay
branching ratio are presented in the context of supersymmetric theories.
Theoretical accuracy increased by inclusion of the next-to-leading order QCD
corrections makes it possible to significantly decrease scale dependence of
the result.
Comparison with the latest CLEO experimental results then allows to limit
supersymmetric loop contributions to the process and consequently constrain the
parameter space of supersymmetric extensions of the
Standard Model. We discuss these constraints both in the minimal supergravity
inspired model (SUGRA) and in the simplest gauge-mediated supersymmetry
breaking model (GMSB) both of which are interesting from the point of view of
the searches for supersymmetry at present and future colliders.
Our analysis also includes the interesting region of large $\tan \beta$ 
relevant for models with Yukawa coupling unification.}

\begin{figure}[t]
\centerline{\psfig{figure=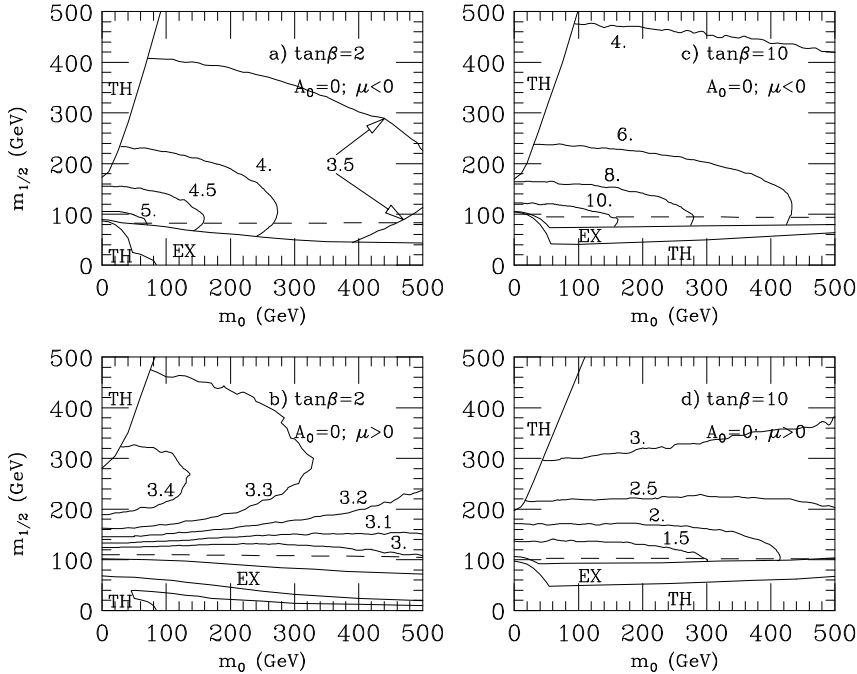,height=3.5in}}
\caption{ Plot of contours of constant branching ratio $B(b\to s\gamma )$ in 
the $m_0\ vs.\ m_{1/2}$
plane, where $A_0=0$ and $m_t=175$ GeV. 
Each contour should be multiplied
by $10^{-4}$. 
The regions labelled by TH (EX) are excluded by theoretical (experimental)
considerations. The dashed contour corresponds to $m_{\tw_1}>80$ GeV for 
a gaugino-like chargino. \label{fig:sugra}}
\end{figure}

An important experimental check of SUSY models is provided by the measurement 
of the $b\to s\gamma$ decay rate. Weak scale SUSY particles contribute to the one loop 
decay amplitude and their presence can significantly modify the Standard Model 
(SM) result.

The best experimental value available for the inclusive $B\to X_s \gamma$ 
branching ratio as measured by the CLEO collaboration $B(B\to X_s 
\gamma)=(2.32\pm0.57 \pm 0.35)\times 10^{-4}$ has to be compared to the 
next-to-leading order Standard Model prediction including non-perturbative 
corrections $B(B\to X_s) \gamma=(3.38\pm 0.33)\times 10^{-4}$, see 
\cite{misiak}. 
SUSY contributions can correct the theoretical value in both directions and a 
requirement that the branching ratio remain restricted within the experimental 
95\% confidence level band $1\times 10^{-4}< B(B\to X_s\gamma )<4.2\times 
10^{-4}$ translates into constraints on the supersymmetric parameter space. All
the next-to-leading order calculation ingredients for SUSY models are already 
available with the exception of the exact two-loop matching conditions between 
the full theory and the effective theory, which have been evaluated only for 
the SM contribution, the charged Higgs contribution and for some special cases 
of the chargino contribution . Here we adopt an approximate procedure 
effectively decoupling the heavy particles in the loop at different scales 
for the dominating SUSY contributions to approximate the exact matching
conditions \cite{bb1}.  

\begin{figure}[t]
\centerline{\psfig{figure=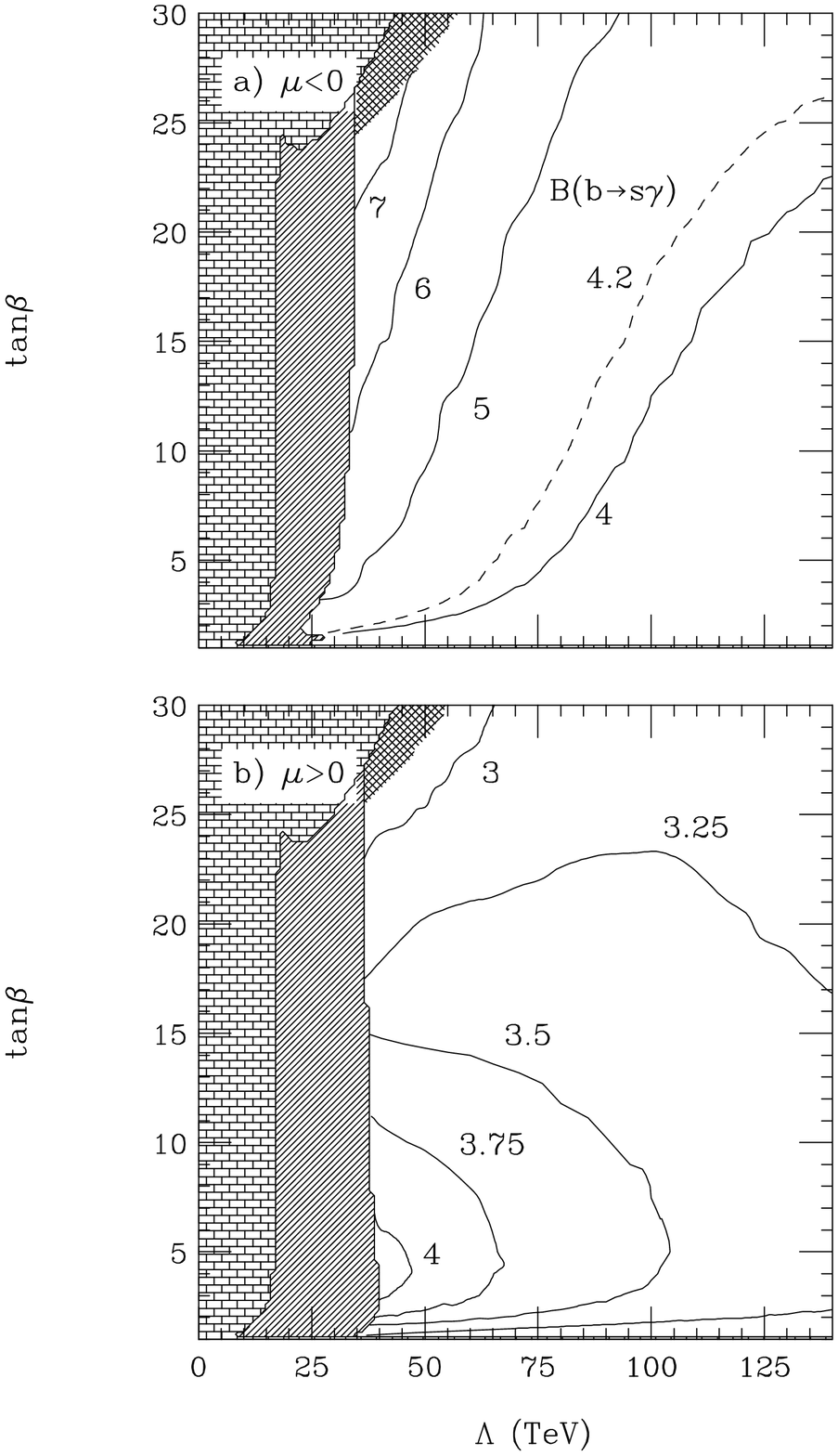,height=3.8in}}
\caption{ Contours of constant branching ratio $B(b\to s\gamma )$ in 
the $\Lambda \ vs.\ \tan\beta$
plane, for $M_{mes}=500 \rm{TeV}$. 
Each contour should be multiplied
by $10^{-4}$. The bricked (shaded) regions are excluded by theoretical 
(experimental) considerations. The cross hatched region is where $\ttau_{1}$ 
is the NLSP. \label{fig:gmsb}}
\end{figure}

In the minimal supersymmetric extension of the SM, the SUSY contribution 
involves the charged Higgs-top quark loop, 
chargino-squark loops, gluino-squark loops and neutralino-squark loops
\cite{masiero}. The first two are most important while the gluino and 
neutralino loops can be safely neglected within the two major classes of SUSY
models, namely the supergravity (SUGRA) inspired models and models with gauge
mediated SUSY breaking (GMSB). The charged Higgs 
loop contribution is always of the same sign as the SM contribution and adds 
constructively with the SM loop. The chargino-squark loops, on the other hand, 
can interfere with the SM contribution either constructively or destructively 
depending on the sign of $\mu$. This feature together with the fact that the 
magnitude of the total chargino-squark contribution grows with $\tan\beta$ 
characterizes constraints imposed on the minimal SUGRA parameter space by 
$b\to s\gamma$.

The minimal version of SUGRA models is parametrized in terms of the common 
scalar mass $m_0$, common gaugino mass $m_{1/2}$ and the universal trilinear
parameter $A_0$, which are all evaluated at the unification scale. Furthermore, 
the value of the ratio of the two Higgs VEV's $\tan\beta$ and the sign of the 
supersymmetric Higgs parameter have to be specified.
In the minimal SUGRA models with $\tan\beta \ltap 5$ the chargino squark 
contribution is small enough to 
permit large allowed regions for both signs of $\mu$ yielding larger branching 
ratios for the case of $\mu <0$.  With $\tan\beta$ increasing, the allowed 
region in the $\mu<0$ case narrows down requiring heavy charginos to suppress 
$\tan\beta$ enhancement in the chargino-squark loops while in the $\mu>0$ case 
cancellation between the chargino loops and and the charged Higgs loop makes it 
possible to obtain branching ratios in some cases very close to the CLEO mean 
value and certainly consistent with the experimental limits over the whole 
range of parameters. As an example, Fig.1 displays contours of constant 
branching ratio $B(b\to s \gamma)$ in the $m_0$ vs $m_{1/2}$ plane for 
$\tan\beta =2$ and $10$, and both signs of $\mu$.

In the large $\tan\beta$ parameter region with $\tan\beta \gtap 35$, the 
$\mu>0$ branch of the parameter set is also constarined by  $b\to s \gamma$ 
since a very large chargino-squark contribution drives the branching ratio 
below the lower experimental value of $1\times 10^{-4}$. The $\mu<0$ case 
requires extremely heavy superpartners ($m_{\tg}\gtap 1500\ \rm{GeV}$, 
$m_{\tq}\gtap 1600\ \rm{GeV}$) in order to satisfy experimental constraints on 
$B(b\to s \gamma)$ and thus imposes severe constraints on models with Yukawa 
coupling unification \cite{ltb}.

Additional dependence on $A_0$ does not change the general picture as 
determined by the value of $\tan\beta$ and the sign of $\mu$ but needs to be 
considered for a detailed analysis of minimal SUGRA models since it can change 
the branching ratio by up to several tens of percent in either direction.

In the simplest GMSB model with a single set of messengers in a $5+\bar{5}$
representation of SU(5) SUSY is broken at the messenger scale $M_{mes}$ and 
both scalar and gaugino masses are proportional to their corresponding
quantum numbers and a single scale $\Lambda$.
Here also the value of $\tan\beta$ and the sign of $\mu$ determine the way
in which the chargino-squark contribution is added to the SM and Higgs 
contributions. Generally, the squarks tend to be significantly heavier than
the charginos in this type of models, which correspondingly suppresses the 
magnitude of the chargino contribution. As a result, for $\mu>0$ the effects 
from chargino and charged Higgs loops more or less cancel each other as they 
interfere destructively ans $b\to s\gamma$ imposes no constraints over much 
of the parameter space as can be seen in Fig.2. In the $\mu<0$ case, the 
$b\to s\gamma$ constraint becomes increasingly restrictive with growing 
$\tan\beta$ since the $\tan\beta$ enhancement of the chargino contribution 
pushes the value of the $b\to s\gamma$ branching ratio over the CLEO experimental 
limit \cite{bbct}.

\end{document}